\documentclass[10pt,draftcls,onecolumn]{IEEEtran}

\IEEEoverridecommandlockouts 

\usepackage[noadjust]{cite}
\usepackage[dvips]{graphicx}
\graphicspath{{./figures/}}

\usepackage[cmex10]{amsmath}
\usepackage{amssymb}
\usepackage{amsfonts}

\usepackage{algorithmic}

\newtheorem{theorem}{Theorem}
\newtheorem{definition}{Definition}
\newtheorem{lemma}{Lemma}

\newtheorem{corollary}{Corollary}

\date{}

\begin{document}
\title{ARQ for Network Coding\thanks{This work is supported by the DARPA ITMANET grant and by the NSF grant CNS-0627021 (NeTS:NBD: XORs in the Air: Practical Wireless Network Coding)}}

\author{Jay~Kumar~Sundararajan, Devavrat~Shah and Muriel~M\'edard\\Laboratory for Information and Decision Systems\\
Massachusetts Institute of Technology, Cambridge, MA 02139, USA.\\Email: \{jaykumar,devavrat,medard\}@mit.edu
}

\maketitle

\begin{abstract}
A new coding and queue management algorithm is proposed for communication networks that employ linear network coding. The algorithm has the feature that the encoding process is truly online, as opposed to a block-by-block approach. The setup assumes a packet erasure broadcast channel with stochastic arrivals and full feedback, but the proposed scheme is potentially applicable to more general lossy networks with link-by-link feedback. The algorithm guarantees that the physical queue size at the sender tracks the backlog in degrees of freedom (also called the virtual queue size). The new notion of a node ``seeing'' a packet is introduced. In terms of this idea, our algorithm may be viewed as a natural extension of ARQ schemes to coded networks. Our approach, known as the drop-when-seen algorithm, is compared with a baseline queuing approach called drop-when-decoded. It is shown that the expected queue size for our approach is $O\left(\frac1{1-\rho}\right)$ as opposed to $\Omega\left(\frac1{(1-\rho)^2}\right)$ for the baseline approach, where $\rho$ is the load factor. 
\end{abstract}

\IEEEpeerreviewmaketitle

\section{Introduction}
Digital fountain codes (\cite{ltcodes,raptor}) are a well-known solution to the problem of reliable communication over a packet erasure channel. 
They have low complexity and require no feedback, except to signal successful decoding of a block.
However, fountain codes do not extend readily to a network setting. Consider a two-link tandem network. If the middle node applies a fountain code on coded packets it has received so far, this does not mean that the overall code from the sender to the receiver will have the properties of a fountain code. In this sense, the fountain code approach is not composable across links. A decode and re-encode scheme will be sub-optimal in terms of delay as pointed out by \cite{linenetworks}. 

In comparison, the random linear network coding solution \cite{desmondallerton}, is composable because it removes the need for decoding at intermediate nodes. 
This solution ensures that with high probability, the transmitted packet will have the \emph{\bf innovation guarantee property}, \emph{i.e.}, it will bring new information
to every receiver, except in the case when the receiver already knows as much as the sender. Thus, every successful reception will bring a unit of new information. This scheme is shown to achieve capacity for the case of a multicast connection. 

However, both fountain codes and random linear network coding have the problem of decoding delay. Both schemes are block-based. While this works for a file download setting, many applications today need to broadcast a continuous stream of packets in real-time. The above schemes would segment the stream into blocks (also called generations) and process one block at a time. If playback can begin only after receiving a full block, then high throughput would require a large delay. 

This raises the interesting question -- can playback begin even before the full block is received? In general, playback is possible till the point up to which all packets have been recovered, which we call the \emph{front of contiguous knowledge}. Traditionally, no incentive is placed on decoding a part of the message using a part of the codeword. In a streaming application however, decoding older packets earlier reduces delay. Performance depends on not only \emph{how much data} is transferred in a given time, but also \emph{which part of the data}. In other words, we are more interested in packet delay than block delay. These issues have been studied by \cite{eminthesis} and \cite{sahai} in a point-to-point setting. However, in a network setting, the problem is not well understood. 
In related work, \cite{sujay}, \cite{rtoblivious} address the question of how many original packets are revealed before the whole block is decoded.
However, playback delay depends on not just the number of recovered packets, but also the order in which they are recovered.

Suppose we have full feedback, then reliable communication over a packet erasure channel can be achieved using Automatic Repeat reQuest (ARQ). This simple scheme achieves 100\% throughput, in-order delivery and the lowest possible packet delay, and it is composable across links. In case of tandem networks, this is the optimal solution. However, link-by-link ARQ cannot achieve the multicast capacity of a general network. The well-known butterfly network is an example. Besides, ARQ does not work well with broadcast-mode links because retransmitting a packet that some receivers did not get is wasteful for the others that already have it. In contrast, network coding readily extends to broadcast-mode links and also achieves multicast capacity of any network. 

Our new scheme combines the benefits of network coding and ARQ by \emph{\bf acknowledging degrees of freedom 
instead of original packets}. (Here, \emph{degree of freedom} refers to a new dimension in the appropriate vector space.) This new framework allows the code to incorporate receivers' states of knowledge and thereby enables the sender to control the evolution of the front of contiguous knowledge. The scheme may thus be viewed as a first step in feedback-based control of the tradeoff between throughput and decoding delay, along the lines suggested in \cite{desmondfeedback}. 

This new kind of feedback is also useful in queue management. Consider a packet erasure broadcast channel. The network coding solution of \cite{desmondallerton} requires the sender to generate a linear combination using potentially all packets of a generation that have arrived so far. If feedback is used only to signal completion of a generation, then the sender will have to store the entire generation till it is decoded. If instead, receivers ACK each packet upon decoding, then the sender can drop packets that all receivers have decoded.

However, even storing all undecoded packets may be suboptimal. Consider a situation where the sender has $n$ packets and all receivers have received ($n-1$) linear combinations: ($\mathbf{p_1}$+$\mathbf{p_2}$), ($\mathbf{p_2}$+$\mathbf{p_3}$), $\ldots,$ ($\mathbf{p_{n-1}}$+$\mathbf{p_n}$). No packet can be decoded by any receiver, so the sender cannot drop any packet. However, the backlog in degrees of freedom is just 1. It would be enough if the sender stores any one of the $\mathbf{p_i}$'s. The degrees of freedom backlog is also called the ``virtual queue'' (\cite{traceyharish, desmondatilla}). We ideally want the physical queue to track the virtual queue. Now, even if the receivers get degrees of freedom at the specified rate, it is not clear when they would decode the original packets. Hence, the drop-when-decoded scheme will not achieve this goal. In this work, we show that we can achieve this goal if we allow ACKs on degrees of freedom.

\subsection{Our contribution and its implications}\label{contrib}
We propose a new online coding and queue update algorithm that uses ACKs on degrees of freedom to guarantee that the physical queue size at the sender will track the backlog in degrees of freedom. We introduce the notion of a node ``seeing'' a message packet, which is defined as follows. (Note: in this work, we treat packets as vectors over a finite field.)

\begin{definition}[Seeing a packet]
A node is said to have \emph{seen} a packet $\mathbf{p}$ if it has enough information to compute a linear combination of the form $(\mathbf{p} + \mathbf{q})$, where $\mathbf{q}$ is itself a linear combination involving only packets that arrived after $\mathbf{p}$ at the sender. (Decoding implies seeing, as we can pick $\mathbf{q}=\mathbf{0}$.)
\end{definition}

Our new scheme is called the \emph{drop-when-seen} algorithm because \emph{a packet is dropped if all receivers have seen it}. The intuition is that if all receivers have seen $\mathbf{p}$, it is enough for the sender's transmissions to involve only packets beyond $\mathbf{p}$. After decoding these packets, the receivers can compute $\mathbf{q}$ and hence obtain $\mathbf{p}$ as well. Therefore, even if the receivers have not decoded $\mathbf{p}$, no information is lost by dropping it. 

Whereas ARQ ACKs a packet upon decoding, our scheme ACKs a packet when it is seen. This proves useful when there is a broadcast constraint. 
We present a deterministic coding scheme that guarantees that the coded packet, if received successfully, would {\bf simultaneously cause each receiver to see its next unseen packet}. We will prove later that seeing a new packet translates to receiving a new degree of freedom. This means, the innovation guarantee property is satisfied and 100\% throughput can be achieved. The example below explains this algorithm for a simple two-receiver case. Section \ref{formal} extends this scheme to more receivers.

\begin{table*}
\centering
{\footnotesize
\begin{tabular}{|c|p{1in}|p{0.9in}|p{0.6in}|p{0.7in}|p{0.54in}|p{0.7in}|p{0.54in}|}
\hline
Time & Sender's queue                  & Transmitted packet	      & Channel state                     & \multicolumn{2}{c|}A         & \multicolumn{2}{c|}B          \\
     &                                 &                          &                                   & {\footnotesize Decoded} & {\footnotesize Seen but not decoded} & {\footnotesize Decoded} & {\footnotesize Seen but not decoded}  \\
\hline
1    & $\mathbf{p_1}$                           & $\mathbf{p_1}$                    & $\rightarrow$ A, $\nrightarrow$ B & $\mathbf{p_1}$                & -     & -                    & -    \\
\hline
2    & $\mathbf{p_1}$, $\mathbf{p_2}$                    & $\mathbf{p_1}\oplus \mathbf{p_2}$          & $\rightarrow$ A, $\rightarrow$  B & $\mathbf{p_1}$, $\mathbf{p_2}$         & -     & -                    & $\mathbf{p_1}$\\
\hline
3    & \ \ \ \ \ $\mathbf{p_2}$, $\mathbf{p_3}$            & $\mathbf{p_2}\oplus \mathbf{p_3}$          & $\nrightarrow$ A, $\rightarrow$  B & $\mathbf{p_1}$, $\mathbf{p_2}$         & -     & -                    & $\mathbf{p_1}, \mathbf{p_2}$\\
\hline
4    & \ \ \ \ \ \ \ \ \ \ $\mathbf{p_3}$         & $\mathbf{p_3}$                    & $\nrightarrow$ A, $\rightarrow$  B & $\mathbf{p_1}$, $\mathbf{p_2}$         & -     & $\mathbf{p_1}, \mathbf{p_2}$, $\mathbf{p_3}$    & - \\
\hline
5    & \ \ \ \ \ \ \ \ \ \ $\mathbf{p_3}$, $\mathbf{p_4}$  & $\mathbf{p_3}\oplus \mathbf{p_4}$          & $\rightarrow$ A, $\nrightarrow$ B & $\mathbf{p_1}, \mathbf{p_2}$           & $\mathbf{p_3}$ & $\mathbf{p_1}, \mathbf{p_2}, \mathbf{p_3}$      & -\\
\hline
6    & \ \ \ \ \ \ \ \ \ \ \ \ \ \ \ $\mathbf{p_4}$ & $\mathbf{p_4}$                    & $\rightarrow$ A, $\rightarrow$  B & $\mathbf{p_1}, \mathbf{p_2}, \mathbf{p_3}, \mathbf{p_4}$ & -     & $\mathbf{p_1}, \mathbf{p_2}, \mathbf{p_3}, \mathbf{p_4}$ & -    \\
\hline
\end{tabular}
}
\caption{An example of the drop-when-seen algorithm}\label{exampletable}
\vspace{-.3in}
\end{table*}

\subsubsection{Example}
Table \ref{exampletable} shows a sample of how the proposed idea works in a packet erasure broadcast channel with two receivers A and B. The sender's queue is shown after the arrival point and before the transmission point of a slot. In each slot, the sender picks the oldest unseen packet for A and B. If they are the same packet, then that packet is sent. If not, their XOR is sent. This rule will cause both receivers to see their oldest unseen packet. In slot 1, $\mathbf{p_1}$ reaches A but not B. In slot 2, $(\mathbf{p_1}\oplus \mathbf{p_2})$ reaches A and B. Since A knows $\mathbf{p_1}$, it can also decode $\mathbf{p_2}$. As for B, it has now seen (but not decoded) $\mathbf{p_1}$. At this point, since A and B have seen $\mathbf{p_1}$, the sender drops it. This is fine because, B will eventually decode $\mathbf{p_2}$ (this happens in slot 4), at which time it can obtain $\mathbf{p_1}$.
Similarly, as shown in the table, $\mathbf{p_2}$, $\mathbf{p_3}$ and $\mathbf{p_4}$ will be dropped in slots 3, 5 and 6 respectively.
However, the drop-when-decoded policy will drop $\mathbf{p_1}$ and $\mathbf{p_2}$ in slot 4, and $\mathbf{p_3}$ and $\mathbf{p_4}$ in slot 6. 
Thus, our new strategy clearly keeps the queue shorter. This is formally proved in Theorem \ref{algm1qsize} and Corollary \ref{algm2qsize}.

\subsubsection{Implications of our new scheme}
\begin{itemize} 
\item 
The information deficit at a receiver is restricted to a window of packets that advances in a streaming manner and has a stable size (namely, the set of unseen packets). In this sense, the proposed encoding scheme is truly online. All receivers see packets in order. 

\item The physical queue size is upper-bounded by the sum of the degrees of freedom backlog between the sender and receivers. Our scheme thus forms a natural bridge between the virtual and physical queue sizes. It can be used to extend results on the stability of virtual queues such as \cite{traceyharish}, \cite{desmondatilla}, \cite{infocom07} to physical queues.

\item At most $n$ packets are involved in the coded packet. This reduces the decoding complexity and the overhead for storing the coding coefficients. 

\item We assume a single packet erasure broadcast channel. But, we believe our algorithm is composable and can be extended to a tandem network of broadcast links, and with suitable modifications, it can be applied to a more general setup like the one in \cite{desmondthesis}.

\item As for decoding delay, we can say that if a receiver receives a packet while it is a leader in terms of number of received degrees of freedom, 
then it will be able to decode all packets up to that point. 
Now, some seen packets might be decoded even before a receiver becomes a leader. The evolution of decoded packets needs further study. 
\end{itemize}

The scheme we proposed in \cite{bergen07} also showed that the physical queue tracks the virtual queues. However, unlike \cite{bergen07} our current work provides an explicit coding scheme that enables us to prove new results. In other work, \cite{keller} also combines feedback and coding to address decoding delay. However, their notion of delay ignores the order in which packets are decoded. Moreover, they do not consider a stream of stochastic arrivals.

\section{The setup}\label{setup}
A sender wants to broadcast a stream of packets to $n$ receivers over a packet erasure broadcast channel. Time is slotted. We focus on linear codes -- every transmission is a linear combination of packets from the incoming stream. 
A node can compute any linear combination whose coefficient vector is in the span of the coefficient vectors of previously received coded packets. This leads to the following definition.

\begin{definition}[Knowledge of a node]
	The \emph{knowledge of a node} is the set of all linear combinations of original packets that it can compute, based on the information it has received so far. The coefficient vectors of these linear combinations form a vector space called the \emph{knowledge space} of the node. 
\end{definition}

The sender has one physical queue with no preset size constraints. We use the notion of a virtual queue to represent the backlog in degrees of freedom between the sender and receiver. There is one virtual queue for each receiver. 
\begin{definition}[Virtual queue]
The size of the $j^{th}$ virtual queue is defined to be the difference between the dimension of the knowledge space of the sender and that of the $j^{th}$ receiver.
\end{definition}

\subsubsection*{Arrivals}
\noindent Packets arrive into the sender's physical queue according to a Bernoulli process 
of rate $\lambda$. 
An arrival at the physical queue translates to an arrival at each virtual queue.

\subsubsection*{Service}
\noindent The channel accepts one packet per slot. Each receiver receives the packet with no errors with probability $\mu$ or an erasure occurs with probability $(1-\mu)$. Erasures occur independently across receivers and across slots. Receivers can detect erasures. 
We assume the innovation guarantee property holds. Then, we can map successful reception to service of the virtual queue. 
Thus, in each slot, every virtual queue is served independently with probability $\mu$. Service of the physical queue will depend on the queue update scheme used. 

\subsubsection*{Feedback}
\noindent In Algorithm 1, feedback is sent when a window of packets is decoded, in order to indicate successful decoding. For Algorithm 2, the feedback is needed in every slot to indicate an erasure. We assume perfect delay-free feedback.

\subsubsection*{Timing}
\begin{figure}
	\centering
		\includegraphics[width=0.45\textwidth]{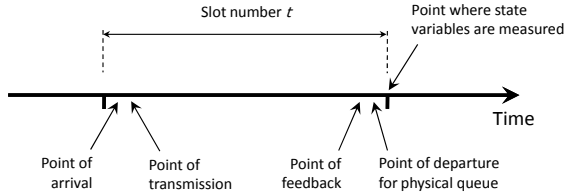}
		\caption{Relative timing of arrival, service and departure points within a slot}\label{earlyarrival}
		\vspace{-.15in}
\end{figure}

\noindent Figure \ref{earlyarrival} shows the relative timing of various events within a slot. 
For simplicity, we assume that the transmission, unless erased by the channel, reaches the receivers before they send feedback for that slot, and feedback from all receivers reaches the sender \emph{before the end of the same slot}. Thus, the feedback incorporates the current slot's reception also. 

\noindent Let $\rho$\ :=\ $\lambda/\mu$. In what follows, we will compare the expected queue size for the baseline drop-when-decoded scheme and the new drop-when-seen scheme, asymptotically as $\rho\rightarrow 1^-$.

\section{Algorithm 1 -- Drop when decoded}\label{algm1section}
The coding scheme assumed is an online version of \cite{desmondallerton}, with no preset generation size. The sender transmits a random linear combination of all packets currently in the queue.  
For any receiver, packets at the sender are unknowns, and each received packet is an equation in these unknowns. Decoding can happen whenever the difference between the number of equations and unknowns involved becomes zero. This difference is essentially the backlog in degrees of freedom, \emph{i.e.}, the virtual queue size. Thus, \emph{successful decoding at a receiver happens when the corresponding virtual queue becomes empty}\footnote{It may be possible to find some unknowns even before the virtual queue becomes empty. 
However, this is a higher order effect and we ignore it.}. 
Whenever a receiver is able to decode in this manner, it informs the sender. Based on this, the sender tracks which receivers have decoded each packet, and drops a packet if it has been decoded by all receivers. 
We assume the field size is large enough to ignore the probability that the coded packet is not innovative. We will now study the behavior of the virtual queues in steady state. But first, we introduce some notation:

\ $Q(t)$\ \ := Size of the physical queue at the end of slot $t$

\ $Q_j(t)$\ := Size of the $j^{th}$ virtual queue at the end of slot $t$

\begin{figure}
\centering
		\includegraphics[width=0.39\textwidth]{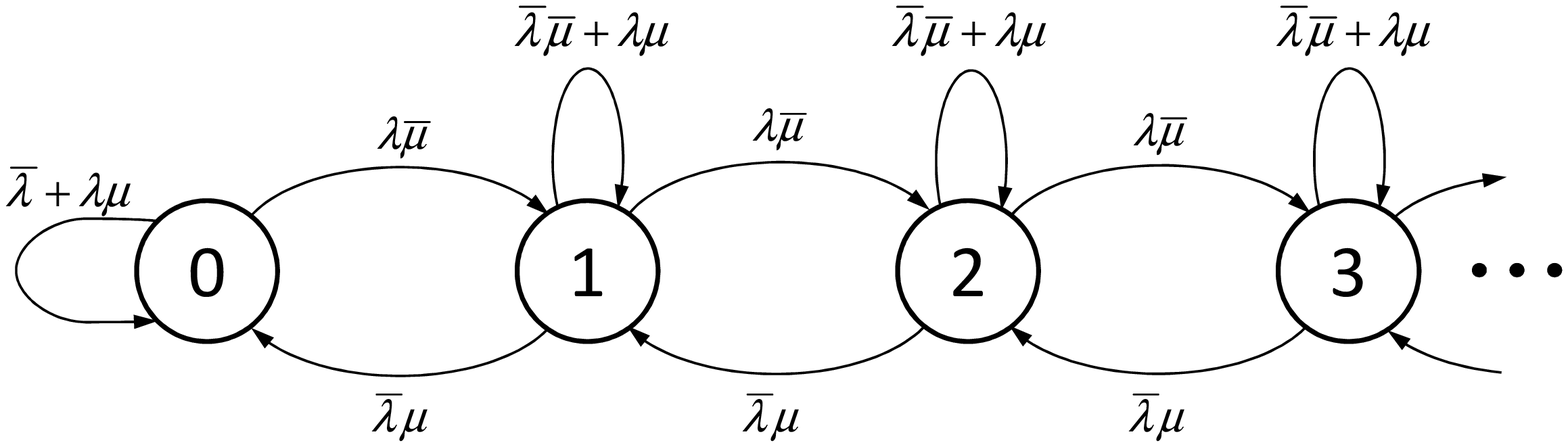}
		\caption{Markov chain for the size of a virtual queue. $\bar{\lambda}$\ :=\ $1-\lambda$;\  $\bar{\mu}$\ :=\ $1-\mu$ }\label{markovchain}
		\vspace{-.23in}
\end{figure}

\noindent Figure \ref{markovchain} shows the Markov chain for $Q_j(t)$. 
If $\rho<1$, the chain is positive recurrent. Its steady state distribution is given by \cite{hunterbook}: \ $\pi_k=(1-\alpha)\alpha^k, k\ge 0$,
where $\alpha=\frac{\lambda(1-\mu)}{\mu(1-\lambda)}$. Thus, the steady state expected size of any virtual queue is:
\begin{equation}
	\lim_{t\rightarrow\infty} \mathbb{E}[Q_j(t)]=\sum_{j=0}^\infty j\pi_j = (1-\mu)\cdot\frac{\rho}{(1-\rho)}
	\label{vqsize}
\end{equation}
Next, we analyze the physical queue size under this scheme.
Let $T$ be the time an arbitrary arrival in steady state spends in the physical queue before departure, excluding the slot in which the arrival occurs.
The packet will not depart until each virtual queue has become empty at least once since its arrival. Let $D_j$ be the time until the next emptying of the $j^{th}$ virtual queue after the new arrival. Then, $T=\max_j D_j$ and so, $\mathbb{E}[T]\ge \mathbb{E}[D_j]$. Hence, we focus on $\mathbb{E}[D_j]$.

We condition on the new arrival seeing state $Q_j=k$ before joining the queue. Then, the state at the end of the slot in which the packet arrives, is $k$ if the channel is ON in that slot and $(k+1)$ otherwise. Now, $D_j$ is simply the first passage time from the state at the end of that slot to state 0. It can be shown that the expected first passage time from state $u$ to state 0 for $u>0$ is given by $\Gamma_{u,0}=\frac{u}{\mu-\lambda}$.  
Now, due to the property that Bernoulli arrivals see time averages (BASTA) \cite{takagi}, the arrival sees the same distribution for $Q_j$ as the steady state distribution given above. 
We can then find $\mathbb{E}[D_j]$ thus:
\begin{eqnarray*}
	\mathbb{E}[D_j]=\sum_{k=0}^\infty \mathbb{P}(\mbox{New arrival sees state $k$}) \mathbb{E}[D_j|\mbox{State $k$}]&&\\
	\ \ \ \ \ =\ \sum_{k=0}^\infty \pi_k \left[\mu\Gamma_{k,0}+(1-\mu)\Gamma_{k+1,0}\right]\ = \frac{1-\mu}{\mu}\cdot\frac{\rho}{(1-\rho)^2}&&\\
\end{eqnarray*}
If the chain is positive recurrent ($\rho<1$), we can use Little's law to find steady state expected physical queue size: 
$\mathbb{E}[Q]=\lambda\mathbb{E}[T]\ge \lambda\mathbb{E}[D_j]$. This leads to the following result:
\begin{theorem}\label{algm1qsize}
\it The expected size of the physical queue in steady state for Algorithm 1 is $\Omega\left(\frac1{(1-\rho)^2}\right)$.
\end{theorem}

\section{Algorithm 2 -- Drop when seen}\label{algm2section}
Algorithm 2 was briefly introduced in Section \ref{contrib}. 
The algorithm uses the notion of reduced row echelon form (RREF) of a matrix in representing the knowledge of a receiver. The definition and properties of the RREF can be found in \cite{artinbook}. 

Let $V$ be the knowledge space of some receiver. Suppose $m$ packets have arrived at the sender so far. 
Then, $V$ must be a subspace of $\mathbb{F}_q^m$ and can be represented using a $dim(V)\times m$ matrix over $\mathbb{F}_q$ whose rows form a basis of $V$.  Multiple representations are possible depending on the basis chosen. However, if we insist that the matrix must be in RREF, we get a unique representation. This unique RREF basis can be obtained by performing a Gaussian elimination on any other basis matrix. 
In the RREF basis, the first nonzero entry of any row is called a \emph{pivot}. Any column with a pivot is called a \emph{pivot column}. 
The number of pivot columns equals the number of nonzero rows, which is $dim[V]$.
The $k^{th}$ packet to have arrived at the sender is said to have an \emph{index} $k$ and is denoted $\mathbf{p_k}$. The columns are ordered so that column $k$ maps to packet $\mathbf{p_k}$. The next theorem connects the seen packets and the RREF basis. 

\begin{theorem}\label{rreftheorem}
	\it A node has seen a packet with index $k$ if and only if the $k^{th}$ column of the RREF basis $B$ of the knowledge space $V$ of the node is a pivot column. 
\end{theorem}

\IEEEproof
The `if' part is clear. For the `only if', suppose column $k$ of $B$ does not contain a pivot. In any linear combination of the rows, rows with pivot after column $k$ cannot contribute anything to column $k$. Rows with pivot before column $k$ will result in a nonzero term in some column to the left of $k$. So the first nonzero term of any vector in $V$ cannot be in column $k$, \emph{i.e.}, $\mathbf{p_k}$ could not have been seen. 
\endproof

\begin{corollary}\label{witness}
	\it If receiver $j$ has seen packet $\mathbf{p_k}$, then it knows exactly one linear combination of the form $\mathbf{p_k}+\mathbf{q}$ such that $\mathbf{q}$ involves only \emph{\textbf{unseen}} packets with index more than $k$. 
\end{corollary}
\begin{corollary}\label{numberseen}
	\it The number of packets seen by a receiver is equal to the dimension of its knowledge space.
\end{corollary}

\begin{definition}[Witness]
	We denote the unique linear combination guaranteed by Corollary \ref{witness} as $\mathbf{W_j}(\mathbf{p_k})$, the \emph{witness for receiver $j$ seeing $\mathbf{p_k}$}.
\end{definition}

The central idea of the new algorithm is to keep track of seen packets instead of decoded packets. After each slot, every receiver informs the sender whether an erasure occurred, using perfect feedback. 
The aim is to use this feedback to ensure the sender stores just enough data to be able to satisfy the innovation guarantee property. The two main parts of the algorithm are the coding and queue update modules.  

 The coding module computes a linear combination $\mathbf{g}$, which will cause any receiver that receives it, to see its oldest unseen packet. First, the sender computes each receiver's knowledge space using feedback and finds its oldest unseen packet. Only these packets will be involved in $\mathbf{g}$, and hence we call them the \emph{transmit set}. 
A receiver can cancel packets involved in $\mathbf{g}$ that it has already seen, by subtracting suitable multiples of the corresponding witnesses. Therefore, the coefficients for $\mathbf{g}$ should be picked such that for each receiver, after canceling the seen packets, the remaining coefficient of the oldest unseen packet is non-zero. Theorem \ref{correctness} proves that this is possible if the field size is at least $n$. With two receivers, the coding module is a simple XOR based scheme (see Table \ref{exampletable}). 
The coding module readily implies the following queue update rule. \textbf{Drop a packet if all receivers have seen it, since the coding module will not use it ever again.} 
Also, while computing the knowledge spaces of the receivers, the sender only needs to track the projection of these spaces on dimensions currently in the queue. 
Thus, the algorithm can be implemented in an incremental manner and the complexity 
tracks the queue size.

\subsection{The formal description of the algorithm}\label{formal}

\subsubsection*{The drop-when-seen algorithm}
\noindent The algorithm works with the RREF bases of the receivers' knowledge spaces. The representation is in the form of coefficient vectors in terms of the current queue contents and not the original packet stream.

\begin{enumerate}
\item Initialize matrices $B_1, \ldots, B_n$ to the empty matrix. 

\item {\it Incorporate new arrivals:}
 Suppose there are $a$ new arrivals. Add the new packets to the end of the queue. 
Append $a$ zeros to every row in each $B_j$. 

\item {\it Transmission: }
If the queue is empty, do nothing; else compute $\mathbf{g}$ using the coding module 
and transmit it.

\item {\it Incorporate channel state feedback: }

For every receiver $j=1$ to $n$, do:

If receiver $j$ received the transmission, include the coefficient vector of $\mathbf{g}$ in terms of the current queue contents, as a new row in $B_j$. Perform Gaussian elimination.

\item {\it Separate out packets that all receivers have seen: } 

Update the following sets and bases:

\ \ $S_j$\ \ := Set of indices of pivot columns of $B_j$

\ \ $S_{\Delta}$\ := $\cap_{j=1}^n S_j$ (set of packets seen by all receivers).

New $B_j$\ := Sub-matrix of current $B_j$ obtained by excluding columns in $S_\Delta$ and corresponding pivot rows.

\item {\it Update the queue: }
Drop the packets with indices in $S_\Delta$. 

\item Go back to step 2 for the next slot.
\end{enumerate}

\noindent {\it The coding module:} 

\noindent Let $\{u_1, u_2, \ldots , u_m\}$ be the set of indices of the oldest unseen packets of the receivers, sorted in ascending order ($m$$\le$$n$, since the oldest unseen packet may be the same for some receivers). Exclude receivers whose oldest unseen packets have not yet arrived at the sender.
Let $R(u_i)$ be the set of receivers whose oldest unseen packet is $\mathbf{p_{u_i}}$.
We now present the coding module to select the linear combination for transmission.

\begin{enumerate}
	\item {\it Loop over oldest unseen packets}
		
		For $j=1$ to $m$, do:

			All receivers in $R(u_j)$ have seen packets $\mathbf{p_{u_i}}$ for $i<j$. Now, $\forall r \in R(u_j)$, find $\mathbf{y_r}:=\sum_{i=1}^{j-1} \alpha_i \mathbf{W_r}(\mathbf{p_{u_i}})$, where $\mathbf{W_r}(\mathbf{p_{u_i}})$ is the witness for receiver $r$ seeing $\mathbf{p_{u_i}}$. Pick $\alpha_j \in \mathbb{F}_q$ such that $\alpha_j$ is different from the coefficient of $\mathbf{p_{u_j}}$ in $\mathbf{y_r}$ for each $r\in R(u_j)$. 
	\item {\it Compute the transmit packet: } \ \ 
		$\mathbf{g}:= \sum_{i=1}^m \alpha_i \mathbf{p_{u_i}}$ 
\end{enumerate}

\begin{theorem}\label{correctness}
	\it If the field size is at least $n$, then the coding module picks a linear combination that will cause any receiver to see its oldest unseen packet upon successful reception.
\end{theorem}

\IEEEproof
First we show that a suitable choice always exists for $\alpha_j$. For $r\in R(u_1)$, $\mathbf{y_r}=\mathbf{0}$. So pick $\alpha_1=1$. For $j>1$, $|R(u_j)|$$\le$$(n-1)$. Even if each $\mathbf{y_r}$ for $r\in R(u_j)$ has a different coefficient for $\mathbf{p_{u_j}}$, that covers only $(n-1)$ different field elements. If $q\ge n$, then there is a choice left in $\mathbb{F}_q$ for $\alpha_j$. 
$\forall j$, $\forall r \in R(u_j)$, receiver $r$ knows $\mathbf{y_r}$. Now, $\mathbf{g}$ and $\mathbf{y_r}$ have the same coefficient for all packets with index less than $u_j$, and a different coefficient for $\mathbf{p_{u_j}}$. Hence, $\mathbf{g}-\mathbf{y_r}$ will involve $\mathbf{p_{u_j}}$ and only packets with index beyond $u_j$. So $r$ can see $\mathbf{p_{u_j}}$. 
\endproof
Theorem \ref{rreftheorem} implies that seeing an unseen packet corresponds to receiving an unknown degree of freedom. Hence, Theorem \ref{correctness} essentially says that the innovation guarantee property is satisfied and hence the scheme is throughput optimal. 
\subsection{Connecting the physical and virtual queue sizes}\label{proofsection}
We will need the following notation:

\noindent $S(t)$\ 	\ := Set of packets arrived at sender till the end of slot $t$

\noindent $V(t)$\ := Sender's knowledge space after incorporating the arrivals in slot $t$. This is simply equal to $\mathbb{F}_q^{|S(t)|}$

\noindent $V_j(t)$\ := Receiver $j$'s knowledge space at the end of slot $t$

\noindent $S_j(t)$\ := Set of packets receiver $j$ has seen till end of slot $t$

\begin{lemma}\label{modularitylemma}
\it For $S_1, S_2, \ldots, S_k$ ($k\ge 1$), subsets of a set $S$:
	\begin{equation}\label{connectingineq}
		|S|-|\cap_{i=1}^{k} S_i|\le \sum_{i=1}^k(|S|-|S_i|)
	\end{equation}
\end{lemma}
We omit the proof. We apply this lemma on the sets $S(t)$ and $S_j(t)$, $j$=1 to $n$. Since the queue holds packets not seen by all receivers, $Q(t)=|S(t)|-|\cap_{j=1}^n S_j(t)|$. 
Also, from Corollary \ref{numberseen}, $|S_j(t)|=dim[V_j(t)]$. 
Hence the RHS of (\ref{connectingineq}) becomes $\sum_{j=1}^n \big[ dim[V(t)] - dim[V_j(t)] \big]$, which is the sum of virtual queue sizes.  
This implies the next theorem.

\begin{theorem}\label{mainthm}
\it For Algorithm 2, the physical queue size at the sender is upper-bounded by the sum of the virtual queue sizes, \emph{i.e.}, the sum of the degrees-of-freedom backlog between the sender and the receivers.
\end{theorem}

Theorem \ref{mainthm} and the result in (\ref{vqsize}) lead to this corollary.
\begin{corollary}\label{algm2qsize}
\it	The expected size of the physical queue in steady state for Algorithm 2 is $O\left(\frac1{1-\rho}\right)$.
\end{corollary}

\section{Conclusions and Extensions}\label{conc}
Comparing the results in Theorem \ref{algm1qsize} and Corollary \ref{algm2qsize}, we see that the queue size for the new Algorithm 2 is significantly lower than Algorithm 1. This will prove useful in reducing congestion.
The new algorithm allows the physical queue size to track the virtual queue size. This extends stability and other queuing-theoretic results on virtual queues to physical queues.

We believe the proposed scheme will be robust to delayed or imperfect feedback, just like conventional ARQ. The scheme readily extends to a tandem network of broadcast links (with no mergers) if the intermediate nodes use the evidence packets in place of the original packets. We expect that it will also extend to other topologies with suitable modifications.

We have proposed a natural extension of ARQ for coded networks. This is the first step towards the goal of using feedback on degrees of freedom to control the tradeoff between throughput and decoding delay, by dynamically adjusting the extent to which packets are mixed in the network.

\bibliographystyle{IEEEtran}
\bibliography{References}

\end{document}